\documentclass[reprint, showpacs,
preprintnumbers,
amsmath,amssymb,
aps, pra, floatfix,
superscriptaddress, nofootinbib]{revtex4-1}

\usepackage{caption}
\usepackage{subcaption}
\expandafter\def\csname ver@subfig.sty\endcsname{}

\usepackage{graphicx}% Include figure files
\usepackage{dcolumn}% Align table columns on decimal point
\usepackage{bm}% bold math
\usepackage{hyperref}% add hypertext capabilities
\usepackage[mathlines]{lineno}% Enable numbering of text and display math
\usepackage{xcolor}
\usepackage{import}
\usepackage{lipsum}
\usepackage{booktabs}
\usepackage{times}      % Loads the Times-Roman Fonts
\usepackage{mathptmx}   % Loads the Times-Roman Math Fonts

\usepackage[capitalise, nameinlink]{cleveref}
\usepackage{makecell}
\usepackage{apptools}
\usepackage{subfig}

\begin{document}
\title{Probing the Scalar WIMP-Pion Coupling with the first LUX-ZEPLIN data}

% 1 
\author{J.~Aalbers}
\affiliation{SLAC National Accelerator Laboratory, Menlo Park, CA 94025-7015, USA}
\affiliation{Kavli Institute for Particle Astrophysics and Cosmology, Stanford University, Stanford, CA  94305-4085 USA}

% 2 
\author{D.S.~Akerib}
\affiliation{SLAC National Accelerator Laboratory, Menlo Park, CA 94025-7015, USA}
\affiliation{Kavli Institute for Particle Astrophysics and Cosmology, Stanford University, Stanford, CA  94305-4085 USA}

% 3 
\author{A.K.~Al Musalhi}
\affiliation{University College London (UCL), Department of Physics and Astronomy, London WC1E 6BT, UK}

% 4 
\author{F.~Alder}
\affiliation{University College London (UCL), Department of Physics and Astronomy, London WC1E 6BT, UK}

% 5 
\author{C.S.~Amarasinghe}
% 6 
\affiliation{University of California, Santa Barbara, Department of Physics, Santa Barbara, CA 93106-9530, USA}
\affiliation{University of Michigan, Randall Laboratory of Physics, Ann Arbor, MI 48109-1040, USA}

% 7 
\author{A.~Ames}
\affiliation{SLAC National Accelerator Laboratory, Menlo Park, CA 94025-7015, USA}
\affiliation{Kavli Institute for Particle Astrophysics and Cosmology, Stanford University, Stanford, CA  94305-4085 USA}

% 8 
\author{T.J.~Anderson}
\affiliation{SLAC National Accelerator Laboratory, Menlo Park, CA 94025-7015, USA}
\affiliation{Kavli Institute for Particle Astrophysics and Cosmology, Stanford University, Stanford, CA  94305-4085 USA}

% 9 
\author{N.~Angelides}
\affiliation{Imperial College London, Physics Department, Blackett Laboratory, London SW7 2AZ, UK}

% 10 
\author{H.M.~Ara\'{u}jo}
\affiliation{Imperial College London, Physics Department, Blackett Laboratory, London SW7 2AZ, UK}

% 11 
\author{J.E.~Armstrong}
\affiliation{University of Maryland, Department of Physics, College Park, MD 20742-4111, USA}

% 12 
% ...

\collaboration{LZ Collaboration}
\date{\today}

\begin{abstract}
Weakly interacting massive particles (WIMPs) may interact with a virtual pion that is exchanged between nucleons.
This interaction channel is important to consider in models where the spin-independent isoscalar channel is suppressed.
Using data from the first science run of the LUX-ZEPLIN dark matter experiment, containing 60 live days of data in a 5.5~tonne fiducial mass of liquid xenon, we report the results on a search for WIMP-pion interactions.
We observe no significant excess and set an upper limit of $1.5\times10^{-46}$~cm$^2$ at a 90\% confidence level for a WIMP mass of 33~GeV/c$^2$ for this interaction.
\end{abstract}

\maketitle

\section{\label{sec:introduction}Introduction}

Astrophysical and cosmological evidence suggests that roughly 27\% of the energy density of the universe is composed of nonluminous dark matter (DM)~\cite{DMEvidence:doi:10.1146, DMEvidence:ARBEY2021103865, DMEvidence:RevModPhys.90.045002}.
One of the most popular candidates for DM is the weakly interacting massive particle (WIMP). 
Various extensions to the Standard Model (SM) naturally give rise to WIMPs making searches for them well motivated~\cite{WIMP:snowmass2021, WIMP:directdetection}.
Direct dark matter experiments, such as LUX-ZEPLIN (LZ)~\cite{LZ:Experiment_2020}, XENONnT~\cite{XenonNT:WS_2023}, PandaX~\cite{PandaX4T:SI2023}, DEAP-3600~\cite{DEAP3600:2020_EFT}, and DarkSide-50~\cite{DarkSide-50:eft_2020}, focus primarily on searching for spin-independent (SI) and spin-dependent (SD) interactions between SM particles and WIMPs with masses in the range of a few GeV/c$^{2}$ to several TeV/c$^2$.
Recent results from both LZ~\cite{LZ:SR1WS_2022} and XENONnT~\cite{XenonNT:WS_2023} place stringent limits on these interactions.
In addition to the isoscalar SI channel, the SD channel represents one of the subleading non-relativistic effective field theory (NREFT) operators; recent LZ results use the null results to place constraints in the NREFT channels~\cite{LZ:SR1_NREFT_2023}.

Expanding from NREFT, chiral effective field theory (ChEFT) admits a new class of next-to-leading-order (NLO) contributions, called two-body currents.
One of those two-body currents is the WIMP-pion coupling~\cite{Cirigliano:2012pq}. 
WIMP-pion coupling introduces a new structure factor, allowing the study of this channel exclusively.

In this article, we present the search results for the SI WIMP–pion interaction, using data from the first science run of the LZ experiment.
We find no significant evidence of an excess of events above the backgrounds and report an upper limit of $1.5\times10^{-46}$~cm$^2$ on the process for a 33~GeV/c$^2$ WIMP at a 90\% confidence level.

\section{\label{sec:results}Results}

\begin{figure*}[hbt!]
    \centering
    \begin{subfigure}[b]{0.2\textwidth}
         \centering
         \includegraphics[width=\textwidth]{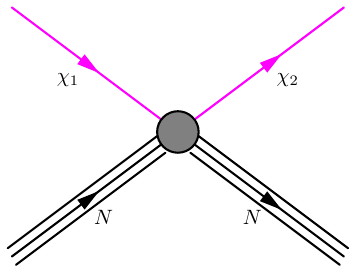}
         \caption{}
         \label{fig:FDs_a}
     \end{subfigure}
     \hfill
     \begin{subfigure}[b]{0.2\textwidth}
     \centering
         \includegraphics[width=\textwidth]{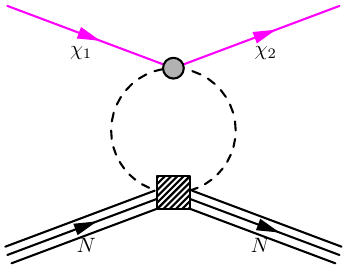}
         \caption{}
         \label{fig:FDs_b}
     \end{subfigure}
     \hfill
     \begin{subfigure}[b]{0.2\textwidth}
     \centering
         \includegraphics[width=\textwidth]{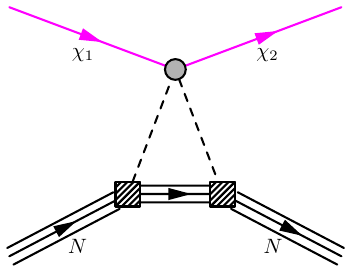}
         \caption{}
         \label{fig:FDs_c}
     \end{subfigure}
     \hfill
     \begin{subfigure}[b]{0.2\textwidth}
     \centering
         \includegraphics[width=\textwidth]{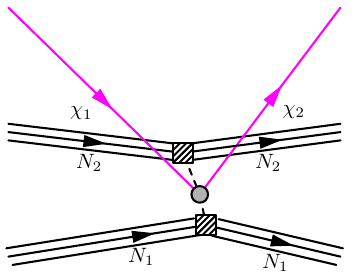}
         \caption{}
         \label{fig:FDs_d}
     \end{subfigure}
    \caption{Feynman diagrams for WIMP-nucleon scattering described by the ChEFT. 
    (a): The LO contribution. 
    The black line represents the nucleon and the magenta line represents the WIMP. 
    The large grey circle denotes the WIMP-nucleon vertex.
    (b,c): The one-loop corrections at NLO. 
    The black dashed line represents the pion loop. 
    The grey circle denotes the WIMP-pion vertex, and the hatched square indicates the pion-nucleon 3-vertex and 4-vertex.
    (d): The two-body current correction at NLO. 
    The dashed line represents the exchanged pion with which the WIMP will interact with.}
    \label{fig:FDs}
\end{figure*}

In a typical analysis of WIMP search data, the WIMP-nucleon scattering is decomposed into the SI channel, where the interaction searched for is assumed to be isoscalar, and the SD channel, where the interaction searched for is with protons or neutrons only.
The SI channel usually emerges as the dominant channel due to its scaling with the square of the number of nucleons~\cite{Engel:1992bf}.
However, neither of these interaction descriptions contain all contributions to their respective channels.
It is therefore beneficial to describe the interaction through the lens of an EFT, where additional contributions can be considered.
When using a ChEFT Lagrangian to describe the scattering of a WIMP with a nucleus (detailed in \autoref{sec:signal_theory}), there is a single diagram at leading order (LO), namely the tree-level $\chi N\rightarrow \chi N$ (\cref{fig:FDs_a}), where $N$ is the nucleon in the nucleus and $\chi$ represents the WIMP.
This LO contribution is included in the standard SI channel. 
At NLO, there are two one-loop diagrams for the single-nucleon process $\chi N\to \chi N$ (\cref{fig:FDs_b,fig:FDs_c}), and one diagram for the two-nucleon process $\chi N_1 N_2\to \chi N_1 N_2$ (\cref{fig:FDs_d}).
In the latter case, the two nucleons exchange a meson (dominantly pions in this energy range), and the WIMP couples to this meson.
From the perspective of ChEFT, the NLO diagrams need to be included in both the SI and SD channels.
Including the two one-loop diagrams change the Wilson coefficients in the standard SI channel whilst the two-body current introduces a new term as a form factor that involves the WIMP-pion vertex.
In the SD channel, only the Wilson coefficients are affected. 
Explicitly, the differential cross-section for the momentum transfer $q=|\vec{q}|$ between the WIMP and the nucleus becomes~\cite{Hoferichter:2016nvd}:
\begin{equation}
\begin{aligned}
     \frac{d\sigma_{\chi N}}{dq^2} &= \frac{1}{4\pi v^2}\left|c_{+} F_{+}(q^2)+c_{-} F_{-}(q^2)+c_{\pi}F_{\pi}(q^2)\right|^2 \\
     & +\frac{1}{v^2(2J+1)}\bigg( |a_{+}|^2S_{00}(q^2)+ Re(a_{+}a_{-}^\star)S_{01}(q^2) \\
     & \qquad\qquad\qquad\qquad +|a_{-}|^2S_{11}(q^2)\bigg)
\end{aligned}
\label{eq:diff_cross_section}
\end{equation}
where the first term represents the SI channel cross-section $\frac{d\sigma_{\chi N}^{\text{SI}}}{dq^2}$, inclusive of the NLO contributions, and the second term corresponds to the SD channel $\frac{d\sigma_{\chi N}^{\text{SD}}}{dq^2}$.
$F$ and $S$ are the form factors for SI and SD scattering, respectively. 
$v$ represents the velocity of the WIMP in the laboratory frame and $J$ denotes the nuclear spin. 
The coefficients $c_i$ and $a_i$ are undetermined constants that will be established through new physics models. 
The subscripts $+/-$ represent the isoscalar/isovector terms, and the subscript $\pi$ signifies the process involving the WIMP-pion coupling.

In \autoref{eq:diff_cross_section}, the SI channel remains dominant over the SD channel because it has the same scaling as the standard SI cross-section, up to some power of $m_\pi/\Lambda_{\text{QCD}}$ and $m_{N}/\Lambda_{\text{QCD}}$. 
Here $m_\pi$ and $ m_{N}$ are the masses for the pion and the nucleon respectively, and $\Lambda_{\text{QCD}}$ is the non-perturbative scale in QCD. 
The SI WIMP-pion coupling is dominant over the SD WIMP-nucleon interaction~\cite{Xenon1T:WIMP-pion_2019}.

In this analysis, we examine the SI WIMP-pion coupling term only (\autoref{fig:FDs_d}), and set all other coefficients to zero.
The WIMP-pion coupling becomes the most relevant in scenarios where the WIMP-nucleon contribution to the SI coupling is suppressed, such as in the minimal supersymmetric standard model~\cite{Crivellin_2015}.
The cross section is then:
\begin{equation}
    \frac{d\sigma_{\chi N}^{SI}}{dq^2} \supset \frac{\left|c_{\pi}\right|^2}{4\pi v^2} \left|F_{\pi}(q^2)\right|^2=\frac{\sigma_{\chi\pi}^{\text{scalar}}}{\mu_\pi^2 v^2} \left|F_{\pi}(q^2)\right|^2,
\end{equation}
where the WIMP-pion cross section is $\sigma_{\chi\pi}^{\text{scalar}}=\frac{\mu_\pi^2}{4\pi}|c_{\pi}|^2$, with $\mu_\pi$ denoting the WIMP-pion reduced mass. 

The differential recoil spectrum for this interaction is then given by~\cite{DM_parameters:LEWIN199687_DM_density}:
\begin{equation}
\label{eq: differential recoil rate}
    \frac{dR}{dE_r}=\frac{2\rho \sigma_{\chi\pi}^{\text{scalar}}}{m_\chi \mu^2_\pi}\left|F_{\pi}(q^2)\right|^2\int_{v_\text{min}}^\infty \frac{f(\vec{v},t)}{v} d^3 v,
\end{equation}
where $E_r$ represents the recoil energy, and $m_\chi$ is the WIMP mass.
$\rho$ is the local dark matter density, $f(\vec{v},t)$ is the velocity distribution of dark matter, and $|F_{\pi}|^2$ is the form factor.

\begin{figure}
    \includegraphics[width=\columnwidth]{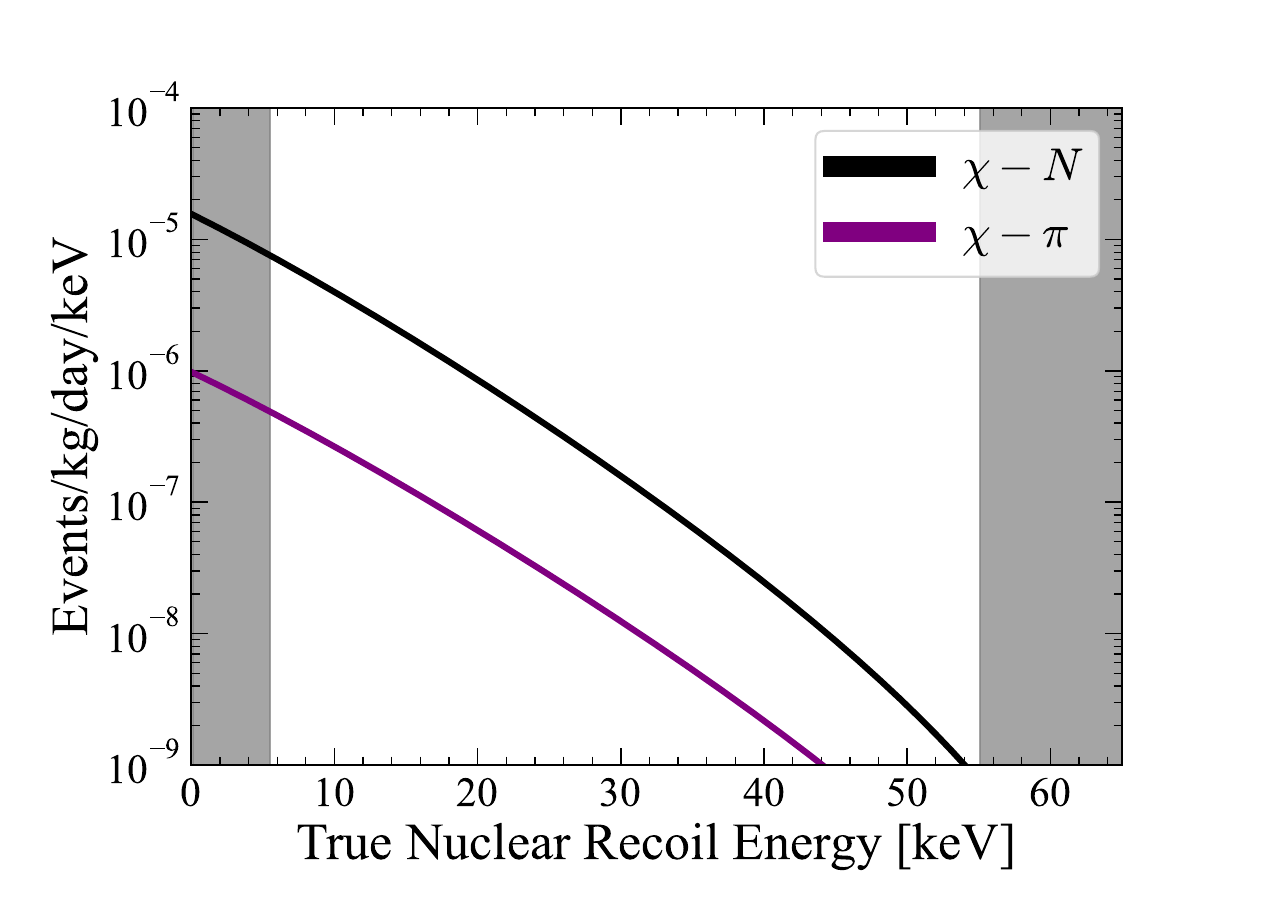}
    \caption{
    Differential recoil spectra for isoscalar SI WIMP-nucleon without any NLO correction (black) and SI WIMP-pion (purple) for a WIMP mass of 30~GeV/$c^2$ on natural xenon.
    The grey regions indicate energy ranges at which signal efficiency falls below 50$\%$.
    The efficiency is assessed using AmLi and tritium calibration data, discussed in \autoref{sec:detector}.
    }
  \label{fig:wimp-pion-recoil}
\end{figure}

To constrain the SI WIMP-pion coupling, data from the first science run (SR1) of the LUX-ZEPLIN (LZ) experiment are used~\cite{LZ:SR1WS_2022}.
The recoil spectra for a SI WIMP-pion interaction with natural xenon, shown in~\autoref{fig:wimp-pion-recoil} along with the spectra from an isoscalar SI WIMP-nucleon interaction without any NLO correction, highlight that both interaction channels produce similar falling exponential spectra.
The similarity in shape between the WIMP-pion and WIMP-nucleon interactions indicates that the same energy window and data selection used in the LZ SR1 SI analysis can be applied here. 

The LZ experiment is a low-background, dual-phase time project chamber (TPC) containing 7~tonnes of liquid xenon (LXe), located in the Davis Campus of the Sanford Underground Research Facility (SURF), South Dakota, USA.  
An energy deposit in the LXe volume produces scintillation photons and ionisation electrons.
The scintillation light from the initial interaction is promptly detected (S1) by an array of photomultiplier tubes.
An electric field, applied across the detector volume, drifts the liberated electrons to the liquid-gas interface.
These electrons are extracted into the xenon gas and are accelerated, producing proportional scintillation and a secondary light signal (S2).
The S1:S2 ratio differs for electronic recoils (ERs) and nuclear recoils (NRs), allowing for discrimination between interaction types.
A correction applied to the S1 and S2 observables (S1$c$ and S2$c$, respectively) accounts for the fact that the response of the detector varies as a function of the vertex position. 
Additional details of the experiment, and run conditions during data collection, can be found in \autoref{sec:detector}. 

The SR1 data, collected between December 2021 and May 2022, contain a total of 60 live days of exposure.
The fiducial volume (FV) used in this analysis is 5.5~$\pm$~0.2~tonnes.
The signal region of interest (ROI) is defined as S1$c$ in the range of 3~--~80~photons detected (phd), S2 greater than 600~phd and S2$c$ less than $10^5$~phd.
A series of selection criteria are applied to the data to remove uncharacteristic detector behaviour and WIMP-like backgrounds, leaving 335 events in the final data set. 
These are described in Ref.~\cite{LZ:SR1WS_2022} along with the data selection efficiency.

\begin{figure}
    \includegraphics[width=\columnwidth]{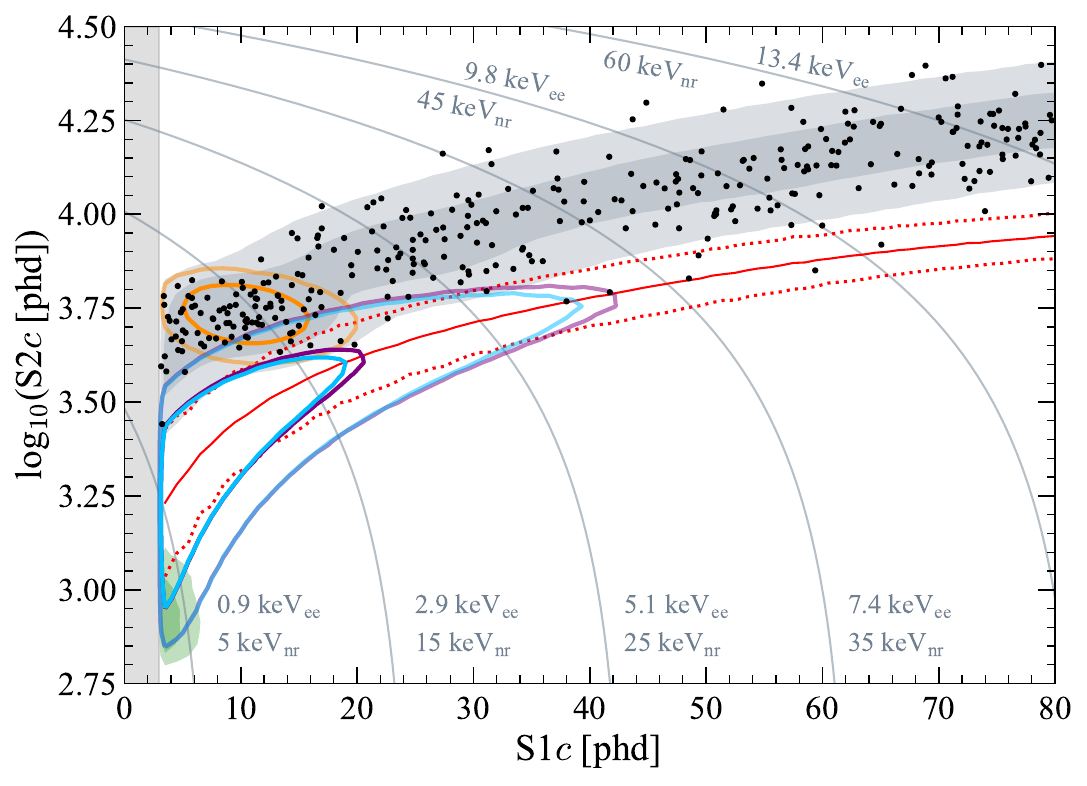}
    \caption{
    The WIMP search data (black points) after all cuts in \{S1$c$, log$_{10}$(S2$c$)\} space.
    The contours that enclose 1$\sigma$ (dark) and 2$\sigma$ (light) regions represent the following background models:
    the shaded grey region indicates the best-fit background model, the orange region indicates ${}^{37}$Ar, and the green region indicates the ${}^{8}$B solar neutrinos. 
    The model for a 30~GeV/c$^2$ WIMP is shown for both an SI WIMP-pion interaction (purple) and an SI WIMP-nucleon interaction (blue).
    The solid red line corresponds to the NR median, while the red dotted lines represent the 10~--~90\% percentiles of the expected response.
    }
  \label{fig:data plot in logS2-S2}
\end{figure}

The expected background contributions in the ROI are given in~\autoref{tab:backgrounds}. 
The ER backgrounds are dominated by radioactive decays in the xenon from ${}^{222}$Rn, ${}^{220}$Rn, ${}^{85}$Kr and ${}^{136}$Xe, as well as interactions from solar neutrinos.
Subdominant backgrounds come from other radioactive impurities in the xenon and detector materials.
The NR backgrounds are from ${}^{8}$B coherent neutrino-nuclear scattering and neutrons emitted from detector materials. 
Events caused by accidental coincidences of unrelated S1 and S2 pulses are also included in the background model.
Detailed compositions of each background are discussed in~\autoref{sec:backgrounds}. 
\autoref{fig:data plot in logS2-S2} shows the data set, as well as backgrounds and the signal distribution for a 30~GeV/c$^2$ WIMP through the SI WIMP-pion interaction.
Also shown is the signal distribution for an interaction through the isoscalar SI WIMP-nucleon channel. 
For any given WIMP mass, an interaction through the WIMP-pion channel will spread over a more extended range in \{S1$c$, log$_{10}$(S2$c$)\} than an interaction through the WIMP-nucleon channel. This can be explained by comparing the recoil spectra in ~\autoref{fig:wimp-pion-recoil}. 
Within the ROI, the gradient of the recoil spectrum of the WIMP-pion coupling is smaller than that of the WIMP-nucleon coupling, causing the WIMP-pion signal to reach the 1$\sigma$ and 2$\sigma$ levels at higher recoil energies.
Although this effect is very small, it may be possible to discriminate between the SI WIMP-nucleon and SI WIMP-pion interactions~\cite{DM:Discriminating_detections_2018}.

\begin{figure}
    \includegraphics[width=\columnwidth]{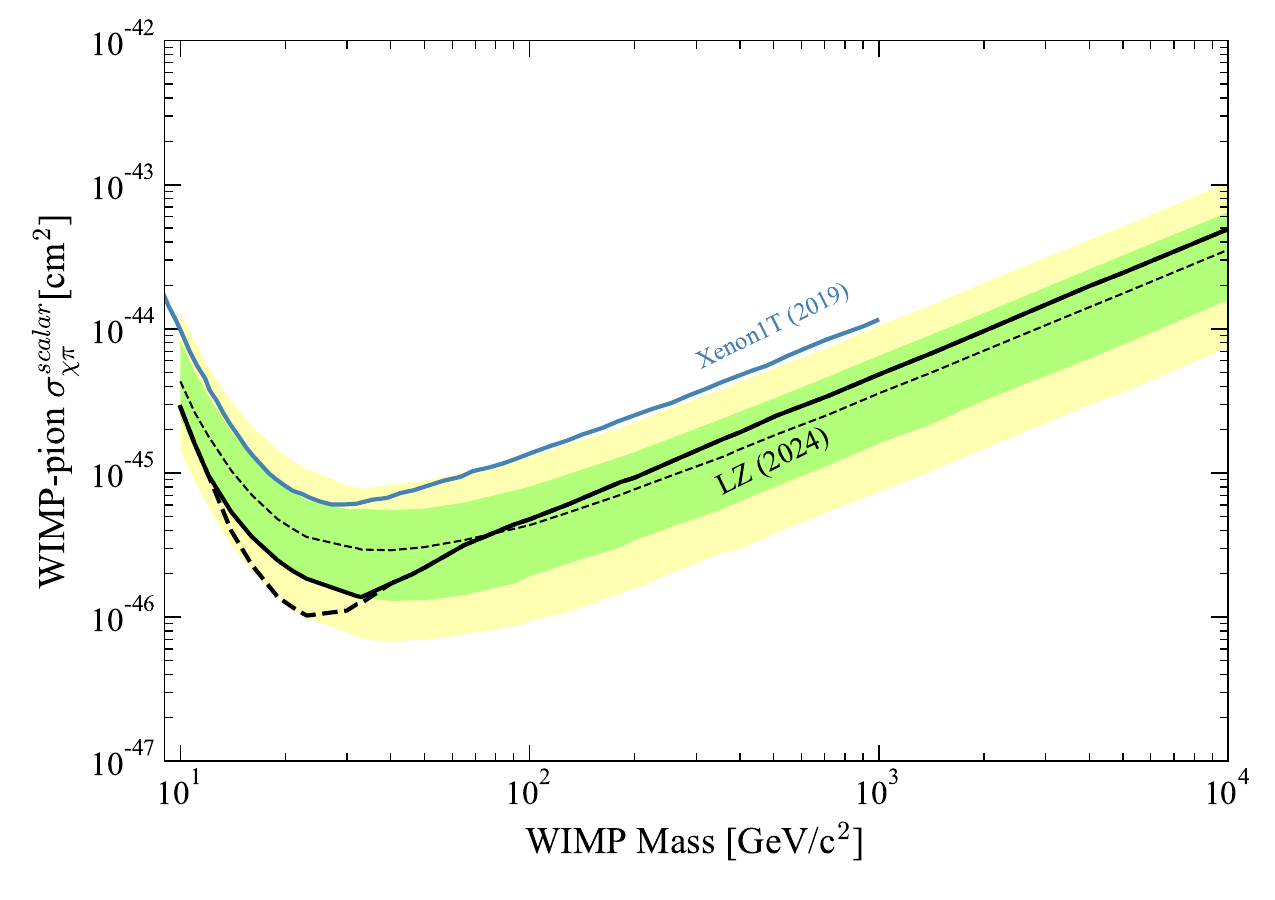}
    \caption{
    The 90\% confidence limit (black lines) of the cross-section for SI WIMP-pion interaction. The black dashed line shows the limit before power constraint.
    The black dotted line shows the median of the sensitivity projection, and the green and yellow bands correspond to the 1$\sigma$ and 2$\sigma$ sensitivity bands, respectively.
    Also shown, in blue, are the WIMP-pion results from XENON1T~\cite{Xenon1T:WIMP-pion_2019}.
    }
  \label{fig:wimp-pion-limit}
\end{figure}

WIMP masses between 9~GeV/c$^2$ and 10,000~GeV/c$^2$ are tested with a frequentist statistical analysis described in \autoref{sec:statistics} and no excess of events above backgrounds are observed at any WIMP mass. 
Shown in \autoref{fig:wimp-pion-limit} are the limits on the SI WIMP-pion interaction cross section, as determined by LZ, together with a previous result from XENON1T~\cite{Xenon1T:WIMP-pion_2019}. 
A power constraint is applied in the 10--20~GeV/c$^2$ mass range to restrict the upper limit from falling 1$\sigma$ below the median expectation from the background-only hypothesis at low energies, such that the probability of excluding a given WIMP-pion cross section if the background-only hypothesis is true does not fall below 16$\%$~\cite{LZ:SR1WS_2022, DM_parameters:BAXTER2021_Conventions, Cowan:2011_power_constraints}. 
The 90\% confidence level upper limit has a minimum of $1.5\times10^{-46}$~cm$^2$ for a 33~GeV/c$^2$ WIMP.

At all masses, the best-fit number of events is 333$\pm$17 events, compared to the 335 in the dataset.
The data are compared to the best-fit background model in a reconstructed energy projection, using an unbinned Kolmogorov-Smirnov test, which results in a p-value of 0.96.
The predicted and best-fit background composition are shown in \autoref{tab:backgrounds}.
The background rates and the p-value are similar to the results in the LZ SI WIMP-nucleon analysis~\cite{LZ:SR1WS_2022}.

\newlength{\dhatheight}
\newcommand{\doublehat}[1]{%
    \settoheight{\dhatheight}{\ensuremath{\hat{#1}}}%
    \addtolength{\dhatheight}{-0.15ex}%
    \hat{\vphantom{\rule{1pt}{\dhatheight}}%
    \smash{\hat{#1}}}}
 
\section{\label{sec:method}Method}

\subsection{\label{sec:detector}Detector description}
The details of the LZ detectors can be found in Ref.~\cite{LZ:Experiment_2020}, and the detector conditions for these data in Ref.~\cite{LZ:SR1WS_2022}.
In this section, the information relevant to this work is summarised.

At the core of the LZ experiment is a cylindrical TPC (described in \autoref{sec:results}), approximately 1.5~m in both height and diameter, containing 7~tonnes of LXe in the active volume.
The TPC is equipped with a total of 494 photomultiplier tubes (PMTs), located at the top and bottom of the TPC to detect scintillation light.
The sensitive volume of the LZ experiment is supported by two veto detectors: a xenon ``Skin'' veto surrounding the active mass, and a near-hermetic ``outer detector'' (OD) consisting of acrylic tanks containing 17~tonnes of gadolinium-loaded liquid scintillator (0.1\% by mass), and surrounded by 238~tonnes of ultra-pure water.
The Skin detector is designed to identify multiple scattering interactions entering or exiting the TPC and is outfitted with 93 1-inch and 38 2-inch PMTs.
The OD is designed to capture and identify neutrons that may scatter in the TPC and is equipped with 120 8-inch PMTs.

The light detected in the TPC (S1 and S2), once position corrected, can be correlated to the number of photons and electrons produced ($n_{ph}$ and $n_e$), by linear scaling factors, $g_1$ and $g_2$:
\begin{equation}
    \text{S1}c = g_1\langle n_{ph} \rangle; \hspace{1cm} \text{S2}c = g_2 \langle n_e \rangle. 
\label{eq:S1S2scale}
\end{equation}
The energy reconstruction of several monoenergetic peaks from background and calibration ER sources are used to determine $g_1$~=~0.114~$\pm$~0.002~phd/photon and $g_2$~=~47.1~$\pm$~1.1~phd/electron.

The ER and NR responses of the TPC detector are measured using dedicated \textit{in-situ} calibrations with tritiated methane (0--18.6~keV ERs) and D-D fusion neutrons (0--74~keV NRs).
The detector and model response parameters from NEST 2.3.7 (Noble Element Simulation Technique)~\cite{NEST:paper_2022, NEST:paper_2023} are tuned to the ER and NR calibration data to reproduce the observed data.
The ER model parameters are propagated into the NR model which is in agreement of better than 1\% for the band medians when compared to DD calibration data.

\subsection{\label{sec:backgrounds}Background composition}

The backgrounds in these data have previously been described~\cite{LZ:SR1Backgrounds_2022} as such only a minimal description is given here.

The ER backgrounds are dominated by radioactive impurities dispersed in the LXe.
These are ${}^{214}$Pb from the ${}^{222}$Rn decay chain, ${}^{212}$Pb from the ${}^{220}$Rn decay chain, and ${}^{85}$Kr.
Peaks in respective decay chains outside of the ROI are used to constrain the rate of ${}^{214}$Pb and ${}^{212}$Pb. 
The concentration of ${}^{85}$Kr is measured with a liquid nitrogen cold trap, and its rate is further validated \textit{in situ} by counting coincident $\beta$-$\gamma$-ray decays~\cite{LZ:SR1Backgrounds_2022}.
All of these components have a near-flat energy spectrum in the ROI and are summed together into the $\beta$ decays background component.
These are considered together with the near-flat contribution from $\gamma$-rays which originate from detector materials and the cavern in which the detector resides~\cite{LZ:radioactivity_and_cleanliness_2020, LZ:cavern_gamma_2020}.
Solar neutrinos, also a near-flat contribution in the ROI, are kept separate as the prediction on the event rate is very precise~\cite{DM_parameters:BAXTER2021_Conventions, Borexino:neutrons_2019, SNO:solar_neutrinos_2013, VINYOLES2017_SolarModels}.
${}^{124}$Xe, a double electron capture, and ${}^{136}$Xe, a double $\beta$ decay, are naturally occurring isotopes, of which the contributions are predicted from known isotopic abundances and decay schemes~\cite{isotopiccompositions_BerglundWieser, Xenon1t:2vec_2019, Exo200:2vbb_2014}. 
Two ER backgrounds are present in this dataset as a result of cosmogenic activation of the xenon prior to arrival at SURF.
Both ${}^{127}$Xe and ${}^{37}$Ar are expected to only contribute significantly to this first science run dataset.
The rate of ${}^{37}$Ar is estimated from the length of time the xenon was exposed to cosmic rays on the surface~\cite{LZ:Ar37_2022}, and ${}^{127}$Xe by K-shell de-excitations and Skin-tagging efficiency for $\gamma$-rays~\cite{LZ:SR1WS_2022}.

The NR backgrounds are comprised of neutrons emitted from radioactive detector materials, through ($\alpha$,n) reactions and spontaneous fission, and from coherent elastic neutrino-nucleus scattering.
The rate of radiogenic neutrons is constrained by the rate of events tagged by the OD veto detector.
The contribution of ${}^{8}$B solar neutrinos is calculated from the known solar flux~\cite{Borexino:neutrons_2019}.
Finally, isolated S1 and S2 pulses can appear within the same event window, appearing as a single-scatter-like event.
The rate of these accidentals is constrained by sideband samples of single-scatter-like events.

Models for these backgrounds in \{S1$_c$,log$_{10}$(S2$_c$)\} are produced using BACCARAT, a package based on GEANT4~\cite{LZ:simulations_2021, ALLISON2016}, together with a detector model, which is fine-tuned using the NEST detector model.
The header file for NEST 2.3.7 that will reproduce the ER and NR response models used in this analysis is available with Ref.~\cite{LZ:SR1WS_2022}.

\begin{table}[ht]
    \centering
    \caption{Expected and fitted numbers of events from the listed sources in the 60~d~$\times$~5.5~t exposure.
    The middle column shows the predicted number of events with uncertainties as described in the text.
    These uncertainties are used as constraints in a combined fit of the background model.
    The fit result is shown in the right column. 
    Values with a fit result of zero are set to have no lower uncertainty.
    }
    %\begin{ruledtabular}
    \begin{tabular}{ccc}
        \hline 
        \hline 
        Source                     & Expected Events      & Fit Result        \\ \hline
        $\beta$ decays + Det. ER   & 215~$\pm$~36         & 222~$\pm$~16  \\
        $\nu$ ER                   & 27.1~$\pm$~1.6       & 27.2~$\pm$~1.6    \\
        ${}^{124}$Xe               & 5.0~$\pm$~1.4        & 5.2~$\pm$~1.4     \\
        ${}^{127}$Xe               & 9.2~$\pm$~0.8        & 9.3~$\pm$~0.8    \\
        ${}^{136}$Xe               & 15.1~$\pm$~2.4       & 15.2~$\pm$~2.4   \\
        ${}^{8}$B CE$\nu$NS        & 0.14~$\pm$~0.01      & 0.14~$\pm$~0.01   \\
        Accidentals                & 1.2~$\pm$~0.3        & 1.2~$\pm$~0.3    \\ \hline
        Subtotal                   & 273~$\pm$~36         & 280~$\pm$~16 \\ \hline
        ${}^{37}$Ar                & [0, 288]             & 52.5~$\pm$~9.4    \\
        Detector neutrons          & 0.0$^{+0.2}$         & 0.0$^{+0.2}$      \\
        30~GeV/c$^2$ WIMP          & -                    & 0.0$^{+0.6}$      \\ \hline
        Total                      & -                    & 333~$\pm$~17      \\
        \hline \hline
    \end{tabular}
    %\end{ruledtabular}
    \label{tab:backgrounds} 
\end{table}

\subsection{\label{sec:signal_theory}ChEFT Lagrangian}
The ChEFT used in this analysis begins as a minimal extension to the SM to accommodate interactions between WIMPs and nucleons. 
This is introduced via the the following Lagrangian~\cite{Cirigliano:2012pq,Hoferichter:2016nvd}:
\begin{equation}
\centering
\label{eq:Lagrangian}
\begin{aligned}
  \mathcal{L}_{\text{full}} &= \mathcal{L}_{\text{QCD}} \\
& +\frac{1}{\Lambda^2}\sum_{q=u,d,s,c,b,t} \Bigl[ \lambda_q^{V}\bar \chi \gamma^\mu \chi \bar q \gamma_\mu q\\
& \qquad\qquad\qquad\quad +\lambda_q^{A} \bar \chi \gamma^\mu \gamma_5 \chi \bar q \gamma_\mu \gamma_5 q \Bigr] \\
& + \frac{1}{\Lambda^3} \sum_{q=u,d,s,c,b,t} \Bigl[m_q \lambda_q^{S} \bar\chi\chi \bar q q \\
& \qquad\qquad\qquad\quad + \lambda_G \frac{\alpha_s}{\pi} \bar\chi\chi\text{Tr}\left(G_{\mu\nu} G^{\mu\nu}\right) \Bigr]
\end{aligned}
\end{equation}
where $\chi$, $q$, $G_{\mu\nu}$ represent the fermionic WIMP, quark, and gauge fields respectively.
$m_q$ is quark mass, $\alpha_s$ is the strong coupling constant, and $\Lambda$ is the new physics scale. 
The first new term is dimension-6 and can be decomposed into a vector interaction and an axial-vector interaction. 
The second new term is dimension-7, corresponding to the scalar interaction. 
The $\lambda_q$ with superscripts is the coupling constant for each interaction. 

The full theory Lagrangian shown in \autoref{eq:Lagrangian} works for any energy scale, including low-energy WIMP-nucleon scattering relevant to LZ WIMP-search, its computation is impractically complicated. 
For convenience, in this analysis we use an EFT for the low-energy dynamics described in Refs.~\cite{Cirigliano:2012pq,Hoferichter:2016nvd}.
This is performed via a two-step matching.
First, the theory is first integrated over the heavy quarks (charm, bottom, and top) to obtain an EFT called Heavy Quark Effective Theory (HQET)~\cite{Georgi:1990um}.
Next, the HQET Lagrangian is matched to the WIMP-nucleon Lagrangian at lower energies. 
This step is done in a ChEFT framework~\cite{Weinberg:1978kz}, and the full derivation can be found in Ref.~\cite{Cirigliano:2012pq}.

As an example, the scalar interaction in \autoref{eq:Lagrangian} can be treated as external sources in ChEFT which are incorporated into the chiral Langragian~\cite{WEINBERG19913}.
For a pure meson theory, $\mathcal{L}_{M}=\mathcal{L}_{M}^{(2)}+\mathcal{L}_{M}^{(4)}+\cdots$, based on the power counting $\mathcal{L}_{M}^{(2n)}\sim p^{2n}$, $m_q\sim \mathcal{O}(p^2)$ and $\lambda_q^{S}\sim \lambda_G \sim \mathcal{O}(p^0)$ where the first term reads as~\cite{Cirigliano:2012pq}:
\begin{equation}
\mathcal{L}_{M}^{(2)}=\frac{F^2}{4}\text{Tr}\left[\partial_{\mu}U^\dagger\partial^{\mu}U\right]+\frac{B_0F^2}{2}\text{Tr}\left[m_q(1-\lambda^S)(U+U^\dagger)\right]
\label{eq:mesonL}
\end{equation}
Here $U=\exp\left(\frac{i}{F}\sum_a t^a \phi_a\right)$ is the meson field with $F$ the pion decay constant and $t^a$ the SU(3) generator. 
The low-energy parameter $B_0$ is defined through  $\langle \bar q q\rangle=-F^2B_0\left(1+\mathcal{O}(m_q)\right)$. 
$\lambda^S$ is the external source with $\lambda^S=\text{diag}(\lambda_\mu^S, \lambda_d^S, \lambda_s^S)$. 
Similarly, one can formulate the chiral Lagrangian incorporating both meson and baryons as $\mathcal{L}_{MB}=\mathcal{L}_{MB}^{(1)}+\mathcal{L}_{MB}^{(2)}+\cdots$, with power counting $\mathcal{L}_{MB}^{(n)}\sim p^{n}$. 
For more details, see Refs.~\cite{Cirigliano:2012pq,Hoferichter:2016nvd,WEINBERG19913}. 
Given the chiral EFT Lagrangian, we are able to write down the Feynman diagrams in ~\autoref{fig:FDs}.

\subsection{\label{sec:recoil_spectrum}Recoil Spectrum}
In this analysis, recoil spectra are generated using ChiralEFT4DM~\cite{chiralEFT4DM:2016}.
The parameters used in the spectra generation follow the convention set in Ref.~\cite{DM_parameters:BAXTER2021_Conventions}, where the dark matter density $\rho$~=~0.3~GeV/cm$^3$~\cite{DM_parameters:LEWIN199687_DM_density}, and the velocity distribution is described by the Standard Halo Model with $\vec{v}_\circledast$~=~(11.1, 12.2, 7.3)~km/s (solar peculiar velocity)~\cite{Schoenrich:Local_kinematics}, $\vec{v}_0$~=~(0, 238, 0)~km/s (local standard of rest velocity)~\cite{DM_parameters:galaxy_context_rest_velocity_1, DM_parameters:galaxy_context_rest_velocity_2} and $v_{\text{esc}}$~=~544~km/s (galactic escape speed)~\cite{DM_parameters:RAVE_survey_escape_velocity}.
For a zero-momentum scalar SI WIMP-pion cross-section, $\sigma_{\chi\pi,0}^{scalar}$, a value of 10$^{-46}$ cm$^2$ was used; based on constraints from XENON1T~\cite{Xenon1T:WIMP-pion_2019}.
The xenon target is considered as a mixture of xenon isotopes weighted by their natural abundances.
The form factor uses the fit described in Ref.~\cite{Hoferichter:2016nvd} to the shell-model calculations of Refs.~\cite{Vietze:2014vsa,Hoferichter:2018acd,Caurier:2004gf}. 

\subsection{\label{sec:statistics}Statistical Method}
A hypothesis test is performed to evaluate the consistency of the observed data with the presence of WIMPs interacting through the SI WIMP-pion channel.
The test statistic utilised for assessing the Parameter of Interest (POI) is defined as a negative log-likelihood, $q = -2\ln{(\lambda)}$. 
Here, $\lambda$ represents the Profile Likelihood Ratio (PLR), defined as:
\begin{equation}
    \lambda(\mu) = \frac{\mathcal{L}(\mu, \hat{\hat{\theta}})}{\mathcal{L}(\hat{\mu}, \hat{\theta})},
\label{eq:statistical_lambda}
\end{equation}
where $\mu$ is the POI, taken to be the number of WIMP-pion scatterings, and $\theta$ represents a set of nuisance parameters related to the contribution of individual background components to the overall observed number of events.
Variables marked with hats represent parameters that maximise the likelihood globally, while those marked with double hats maximise the likelihood for a fixed POI.

For this analysis, an extended unbinned two-sided likelihood fit to the search data in \{S1$_c$, log$_{10}$(S2$_c$)\} is performed; following the approach in Ref.~\cite{DM_parameters:BAXTER2021_Conventions}. 
The observable space ($x_e$) is then in 2D with \{S1$_c$, log$_{10}$(S2$_c$)\}.

The likelihood function, $\mathcal{L}$, is defined as:
\begin{equation}
\begin{aligned}
    \mathcal{L}(\theta) =& \mathrm{Poisson}(N_0|\mu_\mathrm{tot}) \\ 
    & \times \prod^{N_0}_{e=1}\frac{1}{\mu_\mathrm{tot}}\bigg(\mu_s f_s(x_e) + \sum^{N_b}_{b=1}\mu_b f_b(x_e) \bigg) \\
    & \times \prod^{N_b}_{b=1} g_b(\mu_b|\nu_b).
\end{aligned}
\label{eq:likelihood}
\end{equation}
The function accounts for Poisson statistics of the observed counts ($N_0$) and the expected number of events ($\mu_{\rm{tot}}$); which is the sum of the expected number of signal ($\mu_s$) and background ($\mu_b$) events which is a function of $\theta$. 
In the summation, $N_b$ is the number of probability density functions of the backgrounds.
Signal and background components are incorporated as functions of the observable parameter ($x_e$), via the inclusion of $f_s(x_e)$ and $f_b(x_e)$, respectively. 
$g_b(\mu_b|\nu_b)$ represents constraint functions on the rate of each background, typically Gaussian distributions with widths defined by the uncertainties of the backgrounds.

A probability density function in observable space is produced for each background and signal under test.
To simulate the background and signal components, the BACCARAT package based on GEANT4~\cite{LZ:simulations_2021, ALLISON2016} is utilised, along with a bespoke simulation of the LZ detector response, which is fine-tuned using the NEST detector model discussed in \autoref{sec:detector}. 
As part of this methodology, the uncertainties associated with the background components are included as constraint terms in a combined fit of the background model to the data.
The PLR calculation is executed using the LZStats codebase~\cite{LZ_Ibles_LZStats_Thesis_ref}.

Background fluctuations can lead to downward fluctuations in the observed limit. 
To protect against this, a power constraint is introduced to not allow the limit to exclude models where the rejection power of the alternate hypothesis is less than $\pi_{\text{crit}}=0.16$~\cite{DM_parameters:BAXTER2021_Conventions, Cowan:2011_power_constraints}.
This restricts the observed limit to not fall below 1$\sigma$ from the median expected limit.

\bibliography{references}

%\noindent \textbf{For Reviews only, highlighted references (optional)} Please select 5–-10 key references and provide a single sentence for each, highlighting the significance of the work.

%\section{\label{sec:data-availability}Data Availability}
%The authors declare that the main data supporting the findings of this study are available within the article.
%Extra data are available from the corresponding author upon request.

%\section*{\label{sec:data-availability}Data Availability}
%The authors declare that the main data supporting the findings of this study are available within the article.
%Extra data are available from the corresponding author upon request.

\section*{\label{sec:acknowledgements}Acknowledgements}
The research supporting this work took place in part at the Sanford Underground Research Facility (SURF) in Lead, South Dakota. 
Funding for this work is supported by the U.S. Department of Energy, Office of Science, Office of High Energy Physics under Contract Numbers DE-AC02-05CH11231, DE-SC0020216, DE-SC0012704, DE-SC0010010, DE-AC02-07CH11359, DE-SC0012161, DE-SC0015910, DE-SC0014223, DE-SC0010813, DE-SC0009999, DE-NA0003180, DE-SC0011702, DE-SC0010072, DE-SC0015708, DE-SC0006605, DE-SC0008475, DE-SC0019193, DE-FG02-10ER46709, UW PRJ82AJ, DE-SC0013542, DE-AC02-76SF00515, DE-SC0018982, DE-SC0019066, DE-SC0015535, DE-SC0019319, DE-SC0024225, DE-SC0024114, DE-AC52-07NA27344, \& DOE-SC0012447.
This research was also supported by U.S. National Science Foundation (NSF); the UKRI’s Science \& Technology Facilities Council under award numbers ST/M003744/1, ST/M003655/1, ST/M003639/1, ST/M003604/1, ST/M003779/1, ST/M003469/1, ST/M003981/1, ST/N000250/1, ST/N000269/1, ST/N000242/1, ST/N000331/1, ST/N000447/1, ST/N000277/1, ST/N000285/1, ST/S000801/1, ST/S000828/1, ST/S000739/1, ST/S000879/1, ST/S000933/1, ST/S000844/1, ST/S000747/1, ST/S000666/1, ST/R003181/1, ST/W000547/1, ST/W000636/1, ST/W000490/1; Portuguese Foundation for Science and Technology (FCT) under award numbers PTDC/FIS-PAR/2831/2020; the Institute for Basic Science, Korea (budget number IBS-R016-D1); the Swiss National Science Foundation (SNSF)  under award number 10001549.
This research was supported by the Australian Government through the Australian Research Council Centre of Excellence for Dark Matter Particle Physics under award number CE200100008.
We acknowledge additional support from the STFC Boulby Underground Laboratory in the U.K., the GridPP~\cite{faulkner2005gridpp,britton2009gridpp} and IRIS Collaborations, in particular at Imperial College London and additional support by the University College London (UCL) Cosmoparticle Initiative, and the University of Zurich.
We acknowledge additional support from the Center for the Fundamental Physics of the Universe, Brown University. 
K.T. Lesko acknowledges the support of Brasenose College and Oxford University. 
The LZ Collaboration acknowledges the key contributions of Dr. Sidney Cahn, Yale University, in the production of calibration sources. 
This research used resources of the National Energy Research Scientific Computing Center, a DOE Office of Science User Facility supported by the Office of Science of the U.S. Department of Energy under Contract No. DE-AC02-05CH11231. 
We gratefully acknowledge support from GitLab through its GitLab for Education Program. 
The University of Edinburgh is a charitable body, registered in Scotland, with the registration number SC005336. 
The assistance of SURF and its personnel in providing physical access and general logistical and technical support is acknowledged. 
We acknowledge the South Dakota Governor's office, the South Dakota Community Foundation, the South Dakota State University Foundation, and the University of South Dakota Foundation for use of xenon.
We also acknowledge the University of Alabama for providing xenon.
For the purpose of open access, the authors have applied a Creative Commons Attribution (CC BY) licence to any Author Accepted Manuscript version arising from this submission.

\section*{\label{sec:author-contributions}Author contributions}
This work is the result of the contributions and efforts of all participating Institutes of the LZ Collaboration.
The collaboration has constructed and operated the LZ apparatus and performed the data processing, calibration, and data selections.
Y.~Qie initiated the WIMP-pion studies and performed the calculation of theoretical models.
S.~Eriksen and Y.~Qie performed the data analysis and hypothesis tests.
S.~Eriksen, B.~Boxer and Y.~Qie prepared the paper draft.
The paper was reviewed by A.~Kaboth, A.~Manalaysay, A.~Wang.
All authors approved the final version of the manuscript.

\section*{\label{sec:data-availability}Data Availability}
The data points associated with the limit curve (\autoref{fig:wimp-pion-limit}) can be found at \href{https://tinyurl.com/LZSR1WIMPPION}{https://tinyurl.com/LZSR1WIMPPION}.
Additional data is available from the corresponding author upon request.

\section*{\label{sec:competing-interests}Competing interests}
The authors declare no competing interests.

%\section*{\label{sec:publishers-note}Publisher’s note}
%Springer Nature remains neutral with regard to jurisdictional claims in published maps and institutional affiliations.

%\section*{Supplementary information (optional)}
%If your article requires supplementary information, please include these files for peer-review. Please note that supplementary information will not be edited.

\end{document}